\documentclass[a4paper,11pt]{article}
\usepackage{latexsym}
\usepackage{rotating}
\author{Wen-Xiu~Ma\footnote{Email: mawx@math.cityu.edu.hk} 
~and Maxim
Pavlov\footnote{Email: maxim@math.cityu.edu.hk}\\
Department of Mathematics, City University of Hong Kong,\\
Kowloon, Hong Kong}
\title{Extending Hamiltonian Operators \\
to Get Bi-Hamiltonian Coupled KdV Systems}
\date{\nonumber}

\begin{document}
\maketitle

\newcommand{\eqnsection}{
   \renewcommand{\theequation}{\thesection.\arabic{equation}}
   \makeatletter
   \csname $addtoreset\endcsname
   \makeatother}
\eqnsection

\newtheorem{thm}{Theorem}[section]
\newtheorem{Le}{Lemma}[section]
\newtheorem{defi}{Definition}[section]
\newcommand{\R}{\mbox{\rm I \hspace{-0.9em} R}}

\def\be{\begin{equation}}
\def\ee{\end{equation}}
\def\bea{\begin{eqnarray}}
\def\eea{\end{eqnarray}}
\def\ba{\begin{array}}
\def\ea{\end{array}}
\def\la {\lambda}
\def \part {\partial}
\def \al {\alpha}
\def \de {\delta}


\begin{abstract}

An analysis of extension of Hamiltonian operators 
from lower order to higher order of matrix 
paves a way for 
constructing Hamiltonian pairs which may result in hereditary operators.
Based on a specific choice of Hamiltonian operators of lower order, 
new local bi-Hamiltonian coupled KdV systems are proposed. 
As a consequence of bi-Hamiltonian structure, they all possess 
infinitely many symmetries and infinitely many 
conserved densities. 

\end{abstract}

\section{Introduction}
\setcounter{equation}{0}

Bi-Hamiltonian structure 
has been identified as one of the basic mechanisms supplying 
conservation laws and commuting flows and even 
constructing multi-gap solutions for a given nonlinear system.
From differential geometric point of view, bi-Hamiltonian structure
can lead to an interesting application which yields 
a key to describe integrable surfaces embedded in pseudo-Euclidean
spaces. For example, in many cases local bi-Hamiltonian structure
provides additional information about some nonlocal Hamiltonian
structures closely connected with metrics of constant curvatures.
There exist other applications 
in handling quantization of Poisson brackets, and studying the problem
of instability, for instance, in fluid mechanics.
What we would like to develop in this paper is to establish 
more local bi-Hamiltonian structures which can engender 
hierarchies of coupled KdV systems. 

Let $u$ be a dependent variable $u=(u^1,\cdots,u^q)^T$, where 
$u^i,\ 1\le i\le q,$ 
depend on the spatial variable $x=(x_1,\cdots, x_p)^T$ 
and on the temporal variable $t$. We
use ${\cal A} ^r$ ($r\ge 1$) to denote the space of $r$ dimensional column 
function vectors depending on $u$ itself 
and its derivatives with respect to the spatial variable $x$. 
Sometimes we write ${\cal A}^r(u)$ 
in order to specify the 
dependent variable $u$.

A natural inner product over ${\cal A}^r(u)$ is given by
\be <\alpha ,\beta >=\int_{\R ^p}(\alpha )^T \beta dx,\ \alpha,
\beta \in {\cal A}^r.\ee
Moreover on ${\cal A}^q(u)$ (note that the dimensions of both
a function vector
in ${\cal A}^q(u)$ and the dependent variable $u$ are the same),
a Lie product can be defined as follows
\be [K,S] =\Bigl. \frac {\part }{\part \varepsilon }
(K(u+ \varepsilon S)-S(u+\varepsilon K))
\Bigr|_{\varepsilon=0},\ K,S\in {\cal A}^q.\ee
Two function vectors $K,S\in {\cal A}^q(u)$ 
are called to be commutative if $[K,S]=0$.
A fundamental conception that we need is  
Hamiltonian operators, which is shown in the following definition.

\begin{defi}
A linear skew-symmetric operator
$J(u):{\cal A}^q(u)\to {\cal A}^q(u)$ is called a Hamiltonian operator
or to be Hamiltonian, if 
the Jacobi identity
\be <\alpha ,J'(u)[J(u)\beta ]\gamma >+\textrm{cycle}(\alpha , \beta , 
\gamma )=0
\label{hamopJacobiidentity} \ee
holds for all 
$\alpha , \beta , \gamma \in {\cal A}^q(u) $.
A pair of operators $J(u),M(u):{\cal A}^q(u)\to {\cal A}^q(u) $
is called a Hamiltonian
 pair if $J(u)+c M(u)$ is always Hamiltonian
for any constant $c $.
\end{defi} 

Associated with a given Hamiltonian operator $J(u)$, the 
Poisson bracket is defined to be 
\be \{\tilde H_1,\tilde H_2\}_J=\int (\frac {\delta \tilde
H_1}{\delta u})^TJ
\frac {\delta \tilde H_2}{\delta u}\,dx,\ 
\tilde H_1,\tilde H_2\in \tilde {\cal A},\ee
where $ {\tilde {\cal A}}$ consists of functionals
$\tilde H=\int Hdx ,\ H\in  \cal { A}.$
Two functionals $ \tilde H_1,\tilde H_2$ are called to be 
commutative under the Poisson bracket associated with $J$, if 
$ \{\tilde H_1,\tilde H_2\}_J=0$. 
A Hamiltonian operator $J: {\cal A}^q\to {\cal A}^q $ has a nice property
\be  [J\frac {\delta  \tilde H_1}{\delta u},
J\frac {\delta  \tilde H_2}{\delta u} ]=J\frac {\delta 
\{  \tilde H_1, \tilde H_2\}_J }{\delta u},\ 
 \tilde H_1 ,\tilde H_2\in  \tilde {\cal A}, \ee
which gives rise to an important relation between symmetries and 
conserved densities for a Hamiltonian system with the Hamiltonian 
operator $J$.  

If we have a Hamiltonian pair $J$ and $M$, one of which is invertible,
for example $J$,
then $\Phi :=MJ^{-1}$ is a hereditary operator 
\cite{GelfandD-FAA1979} \cite{FuchssteinerF-PD1981}.
Furthermore if the adjoint operator $\Psi:=\Phi ^\dagger =J^{-1}M$ of 
this hereditary operator $\Phi =MJ^{-1}$ maps a gradient vector $f_0=
\frac{\delta \tilde H_0}{\delta u}$ to another gradient vector $f_1=\Psi f_0=
\frac{\delta \tilde H_1}{\delta u}$, then all vectors 
$\Psi ^nf_0,\ n\ge 0,$ are gradient, i.e. there exist functionals
$\tilde H_i,\ i\ge 2$,
such that $\Psi ^if_0=\frac{\delta \tilde H_i}{\delta u},\ i\ge 2$ (see
 \cite{Magri-JMP1978} \cite{FokasF-LNC1980} 
\cite{Olver-GTM1071986} for more information). 
Then it follows from bi-Hamiltonian structure that
all systems of evolution equations in the hierarchy 
$u_t=\Phi ^n f_0,\ n\ge 0,$ commute with each other,
i.e. $[\Phi ^mf_0,\Phi ^nf_0]=0,\ m,n\ge0,$ and they  
have infinitely many common conserved densities being commutative 
under two Poisson brackets.
Therefore Hamiltonian pairs 
can pave a way for constructing integrable systems.
However the problem still exits and just turn to how to find Hamiltonian pairs.

In this paper, what we want to develop is to propose
a possible way to generate Hamiltonian pairs. We are successful in 
constructing bi-Hamiltonian coupled KdV systems in such a way.
The paper is organized as follows. We first concentrate on the
techniques for extending Hamiltonian operators 
from lower order to higher order of matrix,
motivated by the idea in \cite{Ma-GJ1986} \cite{TuM-JPDE1992} 
\cite{GhoshC-JPSJ1994}.
Then we will go on to analyze a class of Hamiltonian pairs which 
can yield hereditary operators and eventually new bi-Hamiltonian 
coupled KdV systems. A few of concluding remarks are given
in the final section.

\section{Extending Hamiltonian operators}
\setcounter{equation}{0}

Let us specify our dependent variables
\bea && u_k=(u_k^1,\cdots, u_k^q)^T,\  1\le k\le N,\nonumber \\ &&
u=(u_1^T,\cdots,u_N^T)^T=
(u_1^1,\cdots,u_1^q,\cdots, u_N^1,\cdots, u_N^q)^T,\nonumber \eea 
and introduce a condition 
\be J _k'(u_k)=J _l'(u_l), \ 1\le k,l\le N, \label{linearpofbigJ}\ee 
on a set of given Hamiltonian operators
$J _k(u_k):\ {\cal A}^q(u_k)\to {\cal A}^q(u_k),\ 1\le k\le N$.
This condition requires a kind of linearity property of the involved
operators with regard to their dependent variables.
Such sets of Hamiltonian operators $J _k(u_k)$ do exist. For example,
we can choose 
\[J_k(u_k)=-\frac 14 \part ^3+2\part u_k\part ^{-1} + 2u_k,\ \part =
\frac {\part } {\part x},\ 1\le k\le N.\]  

The problem that we want to handle here 
is how to generate Hamiltonian operators
starting from a given set of $J_k(u_k),\ 1\le k\le N$.
A simplest solution is to make a big operator to be
$J(u)=\textrm{diag}(J_1(u_1),\cdots, J_N(u_N))$. What we want to develop
below is to propose more general structure of Hamiltonian operators.
To the end, we introduce the following structure of
candidates for Hamiltonian operators
 \be J (u)=\Bigl(
\sum_{k=1}^Nc_{ij}^kJ _k(u_k) \Bigr)_{N\times N}, \label{formofbigJ}\ee
where $\{c_{ij}^k|\, i,j,k=1,2,\cdots N\}$ is a set of given constants.
Obviously this big operator $J(u)$ may be viewed as a linear operator
\[J(u): {\cal A}^{Nq}(u)={\underbrace{{\cal A}^q(u)\times \cdots \times 
{\cal A}^q(u)}_{N}}\to  {\cal A}^{Nq}(u)=
{\underbrace{{\cal A}^q(u)\times \cdots \times {\cal A}^q(u)}_{N}}\, ,\]
where a vector function of ${\cal A}^{q}(u)$ depends on 
all the dependent variables $u_1,\cdots,u_N$,
not just certain dependent variable $u_k$,
by defining 
\be J\alpha =\Bigl( ((J\alpha )_1)^T,\cdots,((J\alpha )_N)^T \Bigr)^T,\ 
(J\alpha )_i=\sum_{j,k=1}^N c_{ij}^kJ_k(u_k)\alpha _j,\ 1\le i\le N,
\label{concretedefinitionofbigJ}
 \ee  
where $\alpha =(\alpha _1^T,\cdots , \alpha _N^T)^T,\ \alpha _i\in 
 {\cal A}^{q}(u),\,1\le i\le N.$ 
To guarantee the skew-symmetric property of the big operator $J$,
a symmetric condition on $\{c_{ij}^k\}$ 
\be  c_{ij}^k=c_{ji}^k,\ 1\le i,j,k\le N \label{symmetricconditionofc_{ij}^k}
\ee 
suffices.
The following theorem provides us with a sufficient condition
for keeping the Jacobi identity (\ref{hamopJacobiidentity}).

\begin{thm} \label{modelofbigJ}
If all $J _k(u_k):{\cal A}^{q}(u_k)\to {\cal A}^{q}(u_k),
\ 1\le k\le N,$ are Hamiltonian operators having 
the linearity 
condition (\ref{linearpofbigJ}) 
and the constants $c_{ij}^k,\, 1\le i,j,k\le N$, 
satisfy the symmetric condition 
(\ref{symmetricconditionofc_{ij}^k}) and the following coupled condition
\be 
\sum_{k=1}^Nc_{ij}^kc_{kl}^n=\sum_{k=1}^Nc_{li}^kc_{kj}^n,\ 
1\le i,j,l,n\le N,\label{realcondofC_{ij}^k}\ee
then the operator $J(u):{\cal A}^{Nq}(u)\to {\cal A}^{Nq}(u)$ 
defined by (\ref{formofbigJ}) and (\ref{concretedefinitionofbigJ})
is a Hamiltonian operator. 
\end{thm}

\noindent {\bf Proof:}
We only need to prove that $J (u)$ satisfies the Jacobi identity
(\ref{hamopJacobiidentity}), 
because the linearity property and the skew-symmetric property
of $J$ have already been shown.
Noting that ${\cal A}^q(u)$ is composed of column function vectors,
we suppose for $\alpha , \beta , \gamma \in {\cal A}^{Nq}(u)$ that 
\[\alpha =(\alpha _1^T,\cdots , \alpha _N^T)^T,\ 
\beta =(\beta _1^T,\cdots , \beta _N^T)^T,\
\gamma =(\gamma _1^T,\cdots , \gamma _N^T)^T,
\ \alpha _i,\beta _i,\gamma _i\in {\cal A}^{q}(u),\ 1\le i\le N .\]
Moreover we will utilize a convenient 
notation $(X)_i=X_i,\ X_i\in {\cal A}^q(u),\ 1\le i\le N,$ 
when a function vector $X\in {\cal A}^{Nq}(u)$ itself is complicated.

First from the definition (\ref{formofbigJ}), we have
\bea &&
J'(u)[J\beta ]=\Bigl ( \sum_{k=1}^N c_{ij}^kJ_k'(u_k)[(J\beta )_k]\Bigr)
_{N\times N} ,\nonumber \\ &&
(J'(u)[J\beta ]\gamma )_i=\sum_{j,k=1}^N c_{ij}^k J'_k(u_k)
[(J\beta )_k]\gamma _j ,\ 1\le i\le N. \nonumber \eea
Then taking advantage of  
the concrete definition (\ref{concretedefinitionofbigJ}),
the linearity condition (\ref{linearpofbigJ}) and
the coupled condition (\ref{realcondofC_{ij}^k}),
we can make the following computation
\bea &&
<\alpha , J'(u)[J\beta ]\gamma >+\textrm{cycle}(\alpha ,\beta ,\gamma)
\nonumber \\&=& \sum_{i,j,k=1}^N c_{ij}^k<\alpha _i, J'_k(u_k)[(J\beta )_k]
\gamma _j>+\textrm{cycle}(\alpha ,\beta ,\gamma)\nonumber \\
&=&\sum_{i,j,k=1}^N c_{ij}^k<\alpha _i ,
J'_k(u_k) [\sum_{l,n=1}^Nc_{kl}^nJ_n
(u_n)\beta _l  ] \gamma _j>+\textrm{cycle}(\alpha ,\beta ,\gamma)
\nonumber \\
&=&\sum_{i,j,k,l,n=1}^N 
c_{ij}^kc_{kl}^n <\alpha _i , J'_k(u_k) [J_n
(u_n)\beta _l  ] \gamma _j>+\textrm{cycle}(\alpha ,\beta ,\gamma)
\nonumber \\
&=&\sum_{i,j,l,n=1}^N \sum_{k=1}^N (
c_{ij}^kc_{kl}^n ) <\alpha _i, J'_k(u_k) [J_n
(u_n)\beta _l  ] \gamma _j>+\textrm{cycle}(\alpha ,\beta ,\gamma)
\nonumber \\
&=&\sum_{i,j,l,n=1}^N \sum_{k=1}^N (
c_{ij}^kc_{kl}^n ) <\alpha _i ,J'_n(u_n) [J_n
(u_n)\beta _l  ] \gamma _j>+\textrm{cycle}(\alpha ,\beta ,\gamma)
\nonumber \\
&=&\sum_{i,j,l,n=1}^N \sum_{k=1}^N (
c_{ij}^kc_{kl}^n ) <\alpha _i, J'_n(u_n) [J_n
(u_n)\beta _l  ] \gamma _j>
\nonumber \\
&& + \sum_{i,j,l,n=1}^N \sum_{k=1}^N (
c_{ij}^kc_{kl}^n ) <\beta _i ,J'_n(u_n) [J_n
(u_n)\gamma _l  ] \alpha _j>
\nonumber \\&&+
\sum_{i,j,l,n=1}^N \sum_{k=1}^N (
c_{ij}^kc_{kl}^n ) <\gamma _i, J'_n(u_n) [J_n
(u_n)\alpha _l  ] \beta _j>
\nonumber \\
&=&\sum_{i,j,l,n=1}^N \sum_{k=1}^N (
c_{ij}^kc_{kl}^n )
  \Bigl(< \alpha _i, J'_n(u_n) [J_n
(u_n)\beta _l  ]  \gamma _j>
+\textrm{cycle}(\alpha _i,\beta_l ,\gamma_j) \Bigr)
.\nonumber 
\eea   
Recall each $J_i(u_i)$ is Hamiltonian and it follows from the above equality
that the big operator $J$ defined by (\ref{formofbigJ}) and
(\ref{concretedefinitionofbigJ}) satisfies the Jacobi identity 
(\ref{hamopJacobiidentity}). 
The proof is completed. 
$\vrule width 1mm height 3mm depth 0mm$

We remark that if we consider 
the coefficients $\{c_{ij}^k\}$ to be the structural constants of a
finite-dimensional algebra
with a basis $\textrm{e}_1,\,\textrm{e}_2,\,\cdots ,\textrm{e}_N$ as follows
\be  \textrm{e}_i*\textrm{e}_j=\sum_{k=1}^Nc_{ij}^k\textrm{e}_k,\ 1\le i,j
\le N,\ee 
then the symmetric condition (\ref{symmetricconditionofc_{ij}^k})
and the coupled condition (\ref{realcondofC_{ij}^k}) can become 
\be \textrm{e}_i*\textrm{e}_j=\textrm{e}_j*\textrm{e}_i,\ 1\le i,j\le N,
\ee
\be
 (\textrm{e}_i*\textrm{e}_j)*\textrm{e}_k=(\textrm{e}_k*\textrm{e}_i)
*\textrm{e}_j,\ 1\le i,j,k\le N, \ee
respectively. They reflect two specific properties of the related algebra.

Apparently basic scalar Hamiltonian
operators satisfying the linearity condition 
(\ref{linearpofbigJ}) can be the following set
\be J_i(u_i)=
c_i\part ^3 +d_i\part +2u_{ix}+4u_i\part ,\ 1\le i\le N,
 \label{basicchoiceofJ_i(u_i)}\ee 
where $\part =\part /\part x$ and 
$c _i, d_i,\, 1\le i\le N,  $ are arbitrary constants. 
Of course, matrix Hamiltonian operators 
having the linearity condition (\ref{linearpofbigJ}) may be chosen. 
Actually, such sets of 
Hamiltonian operators may be 
presented directly from the above operators by Theorem \ref{modelofbigJ}
or by perturbation around solutions as in Refs. 
\cite{MaF-PLA1996} \cite{MaF-CSF1996}.

In what follows,
we example applications of Theorem \ref{modelofbigJ} to two specific 
choices of $\{c_{ij}^k\}$. The analysis below is quite
similar to the one
made for the extension of hereditary operators in  Ref. \cite{Ma-JPA1998}.     

\noindent {\bf Example 1:}
Let us first choose 
\be c_{ij}^k=\delta _{i+j,k-p},\ee
where  $p$ is an integer and $\delta_{kl}$ denotes the Kronecker symbol.
The corresponding big operator formed by 
(\ref{formofbigJ}) becomes 
\be J(u)=\left [ \ba {cccc}
J _{p+2}(u_{p+2}) &J _{p+3}(u_{p+3}) & \cdots & J _{p+N+1}(u_{p+N+1})
\vspace{4mm}\\
J _{p+3}(u_{p+3}) &
&\begin{turn}{45}\vdots\end{turn}  & J _{p+N+2}(u_{p+N+2})
\vspace{4mm}\\
\vdots &\begin{turn}{45}\vdots\end{turn}  &
\begin{turn}{45}\vdots\end{turn}  & \vdots \vspace{4mm}\\
J _{p+N+1}(u_{p+N+1}) &J _{p+N+2}(u_{p+N+2}) & \cdots & 
J _{p+2N}(u_{p+2N})
\ea \right] ,\label{Jform1}\ee
where we accept that $J_i(u_i)=0$ if $i\le 0$ or $i\ge N+1$. 

For two cases of $ -2N+1\le p\le -N$ and $ -1\le p\le N-2$,
the coupled condition (\ref{realcondofC_{ij}^k}) can be satisfied, 
because we have
\bea && 
\sum_{k=1}^Nc_{ij}^kc_{kl}^n=\sum_{k=1}^Nc_{li}^kc_{kj}^n
=\left \{ \ba {cl} 
1, &\quad \textrm{when} \ n-i-j-l=2p,\vspace{2mm}\\
0, &\quad \textrm{otherwise} .
\ea \right .\nonumber
 \eea 
While proving the above equality,
we should keep in mind that we have
\[ 1\le i+j+p=n-l-p\le N, \ 1\le i+l+p=n-j-p\le N,\]
when $n-i-j-l=2p$.
But for the case of $-N<p<-1$, upon choosing $i=-p+1, \, j=-p-1,\,
n=l=1$,
we obtain
\[ \sum_{k=1}^Nc_{ij}^kc_{kl}^n=1,\ 
\sum_{k=1}^Nc_{li}^kc_{kj}^n=0,\]
and thus the coupled condition (\ref{realcondofC_{ij}^k}) can not be satisfied.

Note that when $p\ge N-1$ or $p\le -2N$, the resulting operators are all 
zero operators.
Therefore among the operators defined by (\ref{Jform1}), we can obtain
only two sets of candidates for Hamiltonian operators
\bea  &&
 J(u)=\left [ \ba {cccc}
0& & & J_{p+N+1}(u_{p+N+1})
\vspace{4mm}\\ & &J_{p+N+1}(u_{p+N+1}) &J_{p+N+2}(u_{p+N+2})
\vspace{4mm}\\
 &\begin{turn}{45}\vdots\end{turn}  &
\begin{turn}{45}\vdots\end{turn}  & \vdots \vspace{4mm}\\
J _{p+N+1}(u_{p+N+1}) &J _{p+N+2}(u_{p+N+2}) & \cdots & 
J _{p+2N}(u_{p+2N})
\ea \right] , \nonumber \\ &&  \qquad \qquad \qquad \qquad \qquad \qquad 
\qquad \qquad \qquad \qquad \qquad \qquad 
\ -2N+1\le p\le -N,
\label{form1concrete1} 
\\ &&
 J(u)=\left [ \ba {cccc}
J _{p+2}(u_{p+2}) &J _{p+3}(u_{p+3}) & \cdots & J _{p+N+1}(u_{p+N+1})
\vspace{4mm}\\
J _{p+3}(u_{p+3}) & 
&\begin{turn}{45}\vdots\end{turn}  & 
\vspace{4mm}\\
\vdots &\begin{turn}{45}\vdots\end{turn}  & &\vspace{4mm}\\
J _{p+N+1}(u_{p+N+1}) & 
& & 0
\ea \right] ,\ -1\le p\le N-2,
\nonumber \\ &&  \qquad \qquad \qquad \qquad \qquad \qquad 
\qquad \qquad \qquad \qquad \qquad \qquad 
\label{form1concrete2}  \eea
where we still accept that $\Phi_i(u_i)=0$ if $i\le 0$ or $i\ge N+1$. 
These two sets of operators can be changed to each other
by a simple transformation
$(u_1,u_2,\cdots,u_N)\leftrightarrow (u_N,u_{N-1},\cdots,u_1)$. 

\noindent {\bf Example 2:} 
Let us second choose 
\be c_{ij}^k=\delta _{kl},\ l=i+j-p\ (\textrm{mod}\, N) ,\ee
where $1\le p\le N$ is fixed and $\delta_{kl}$ 
denotes the Kronecker symbol again. 
In this case, we have 
\be \sum_{k=1}^Nc_{ij}^kc_{kl}^n=\sum_{k=1}^Nc_{li}^kc_{kj}^n
=\left \{ \ba {cl} 
1, &\quad \textrm{when} \ i+j+l-n=2p 
\ (\textrm{mod}\, N),\vspace{2mm}\\
0, &\quad \textrm{otherwise} ,
\ea \right .\nonumber \ee
which allows us to conclude that the coupled condition 
 (\ref{realcondofC_{ij}^k}) holds, indeed.
Therefore we obtain a set of candidates for 
Hamiltonian operators
\be J(u)=\left [ \ba {cccc}
J _{2-p}(u_{2-p}) &J _{3-p}(u_{3-p}) & \cdots & J _{N-p+1}(u_{N-p+1})
\vspace{4mm}\\
J _{3-p}(u_{3-p}) &  &
  & J _{N-p+2}(u_{N-p+2})
\vspace{4mm}\\
\vdots  &
\begin{turn}{45}\vdots\end{turn}
 & \begin{turn}{45}\vdots\end{turn}
& \vdots \vspace{4mm}\\
J _{N-p+1}(u_{N-p+1}) &J _{N-p+2}(u_{N-p+2}) & \cdots & 
J _{2N-p}(u_{2N-p})
\ea \right] ,\label{Jform2}\ee
where we accept $J_i(u_i)=J_j(u_j)$ if $i=j\ 
(\textrm{mod}\,N)$, while determining the operators involved, for example, 
$J_{2-p}(u_{2-p})=J _{N}(u_N)$ when $p=2$.
A special choice of $p=1$ leads to a candidate of Hamiltonian operators
 \be J(u)= \left [ \ba {cccc} J_1(u_1)&J_2(u_2)&\cdots &J_N(u_N)
\vspace{2mm}\\
J_2(u_2) &
& \begin{turn}{45}\vdots\end{turn} 
& J_1(u_1) \vspace{2mm}\\
\vdots &
\begin{turn}{45}\vdots\end{turn} 
 &\begin{turn}{45}\vdots\end{turn} 
 &\vdots \vspace{2mm}\\
J_N(u_N)& J_{1}(u_{1})& \cdots &J_{N-1}(u_{N-1})
 \ea \right ]. \label{thirdcoupledkdvHamiltonianoperator}
\ee  
This operator will be our starting point for constructing new 
integrable systems in next section.

\section{New bi-Hamiltonian coupled KdV systems}
\setcounter{equation}{0}

From now on, we focus on the candidate of Hamiltonian operators 
defined by (\ref{thirdcoupledkdvHamiltonianoperator}).
Let us pick out the Hamiltonian operators formed by 
(\ref{thirdcoupledkdvHamiltonianoperator}) under a choice of 
(\ref{basicchoiceofJ_i(u_i)}).
Then the following Hamiltonian pair can be engendered
\bea && 
J(u)=A\part =\left [ \ba {cccc}
a_1 & a_2 & \cdots & a_N
\vspace{2mm}\\
a_2 &
&\begin{turn}{45}\vdots\end{turn}  & a_1
\vspace{2mm}\\
\vdots &\begin{turn}{45}\vdots\end{turn}  &
\begin{turn}{45}\vdots\end{turn}  & \vdots \vspace{2mm}\\
a_N &a_1 
& \cdots & a_{N-1}
\ea \right]\part  ,\\
&& M(u)=
\left [ \ba {cccc}
M_1(u_1) & M_2(u_2) & \cdots & M_N(u_N)
\vspace{2mm}\\
M_2(u_2) &
&\begin{turn}{45}\vdots\end{turn}  & M_1(u_1)
\vspace{2mm}\\
\vdots &\begin{turn}{45}\vdots\end{turn}  &
\begin{turn}{45}\vdots\end{turn}  & \vdots \vspace{2mm}\\
M_N(u_N) &M_1(u_1) 
& \cdots & M_{N-1}(u_{N-1})
\ea \right] , 
\eea 
where $\part = {\part }/\part x$, $u=(u_1,u_2,\cdots ,u_N)^T$,
$a_i=\textrm{const.},\,1\le i\le N,$  
and the operators $M_i(u_i),\,1\le i\le N,$ are given by 
\be M_i(u_i)=
c_i\part ^3 +d_i\part +2u_{ix}+4u_i\part ,\ c_i,d_i=\textrm{const.},\ 
1\le i\le N,\ee
which are all Hamiltonian.

We assume that the constant matrix $A$ is invertible to
guarantee the invertibility of $J$, and
its inverse matrix is given by 
\be B= \left [ \ba {cccc}
b_1 & b_2 & \cdots & b_N
\vspace{2mm}\\
b_2 &
&\begin{turn}{45}\vdots\end{turn}  & b_1
\vspace{2mm}\\
\vdots &\begin{turn}{45}\vdots\end{turn}  &
\begin{turn}{45}\vdots\end{turn}  & \vdots \vspace{2mm}\\
b_N &b_1 
& \cdots & b_{N-1}
\ea \right], \ee
where $b_i,\ 1\le i\le N,$ can be determined by solving a specific
linear system
\[\left\{\ba {l}   a_1b_1 + a_2b_2+\cdots +a_Nb_N=1,\vspace{2mm}\\
 a_2b_1 + a_3b_2+\cdots + a_Nb_{N-1}+a_1b_N=0,\vspace{2mm}\\
\cdots \cdots \vspace{2mm}\\
a_Nb_1 + a_{1}b_2+\cdots +a_{N-1}b_N=0.
\ea \right. \]
Now the resulting hereditary operator $\Phi=MJ^{-1}$ reads as 
\bea \Phi (u)&= &\left [ \ba {cccc}
M_1(u_1) & M_2(u_2) & \cdots & M_N(u_N)
\vspace{2mm}\\
M_2(u_2) &
&\begin{turn}{45}\vdots\end{turn}  & M_1(u_1)
\vspace{2mm}\\
\vdots &\begin{turn}{45}\vdots\end{turn}  &
\begin{turn}{45}\vdots\end{turn}  & \vdots \vspace{2mm}\\
M_N(u_N) &M_1(u_1) 
& \cdots & M_{N-1}(u_{N-1})
\ea \right]  \left [ \ba {cccc}
b_1 & b_2 & \cdots & b_N
\vspace{2mm}\\
b_2 &
&\begin{turn}{45}\vdots\end{turn}  & b_1
\vspace{2mm}\\
\vdots &\begin{turn}{45}\vdots\end{turn}  &
\begin{turn}{45}\vdots\end{turn}  & \vdots \vspace{2mm}\\
b_N &b_1 
& \cdots & b_{N-1}
\ea \right]\part ^{-1}\nonumber
\\ &=&
\left [ \ba {cccc}
\Phi _1(u_1) & \Phi _2(u_2) & \cdots & \Phi _N(u_N)
\vspace{2mm}\\
\Phi _2(u_2) &
&\begin{turn}{45}\vdots\end{turn}  & \Phi _1(u_1)
\vspace{2mm}\\
\vdots &\begin{turn}{45}\vdots\end{turn}  &
\begin{turn}{45}\vdots\end{turn}  & \vdots \vspace{2mm}\\
\Phi _N(u_N) &\Phi _1(u_1) 
& \cdots & \Phi _{N-1}(u_{N-1})
\ea \right]  \left [ \ba {cccc}
b_1 & b_2 & \cdots & b_N
\vspace{2mm}\\
b_2 &
&\begin{turn}{45}\vdots\end{turn}  & b_1
\vspace{2mm}\\
\vdots &\begin{turn}{45}\vdots\end{turn}  &
\begin{turn}{45}\vdots\end{turn}  & \vdots \vspace{2mm}\\
b_N &b_1 
& \cdots & b_{N-1}
\ea \right]
 \eea
with
\be \Phi _i(u_i)=M_i(u_i)
\part ^{-1}=
c_i\part ^2 +d_i +2u_{ix}\part ^{-1}+4u_i ,\ 1\le i\le N.
\label{phi_iform}\ee   
This hereditary operator can be rewritten as a concise form
\be \Phi (u)=MJ^{-1}=
\left ( \sum_{k=1}^Nb_{k-i+j}\Phi_k(u_k) \right),\ee 
where $b_i=b_j$ if $i\equiv j\, (\textrm{mod}\,N)$.
It is also an example satisfying 
the extension scheme of hereditary operators in Ref. \cite{Ma-JPA1998},
because upon setting $c_{ij}^k=b_{k-i+j}$ we have three equal
sums for all $1\le i,j,l,n\le N$:  
\bea &&
 \sum_{k=1}^Nc_{ij}^kc_{kn}^l=\sum_{k=1}^Nb_{k-i+j}b_{l-k+n}=
\sum_{m=1}^Nb_{m}b_{l+i-j+n-m}, \nonumber \\ &&
\sum_{k=1}^Nc_{ik}^lc_{kj}^n=\sum_{k=1}^Nb_{l-i+k}b_{n-k+j}=
\sum_{m=1}^Nb_{m}b_{l+i-j+n-m}, \nonumber \\ &&
\sum_{k=1}^Nc_{in}^kc_{kj}^l=\sum_{k=1}^Nb_{k-i+n}b_{l-k+j}=
\sum_{m=1}^Nb_{m}b_{l+i-j+n-m}, \nonumber \eea
which are sufficient for $\Phi (u)$ to be hereditary (see \cite{Ma-JPA1998}).

We now turn our attention to investigating the nonlinear systems
which can be resulted from the above Hamiltonian pairs.
Such first system can be the following   
\be u_t=\Phi u_x=MJ^{-1}u_x, \ee
which can be represented as
\be  u_{it}=\sum_{k,j=1}^N b_{k-i+j}(c_ku_{jxxx}+d_ku_{jx}+2u_{kx}u_j+
4u_ku_{jx}),\ 1\le i\le N, \label{firstsystemin3rdckdvhierarchy}\ee
where we again accept $b_i=b_j$ if $i\equiv j\ (\textrm{mod}\,N)$.  
It is easy to find that 
\be  f_0:=J^{-1}u_x=Bu=\frac {\delta \tilde H_0} {\delta u},
 \ \tilde H_0=\int H_0dx,\ H_0=\frac 12 u^TBu.\ee 
Go ahead to check whether or not 
the next vector defined by 
 \be  f_1:=\Psi f_0=\Phi ^\dagger f_0
= B\left [
\ba {cccc}
\Psi_1(u_1) & \Psi _2(u_2) & \cdots & \Psi _N(u_N)
\vspace{2mm}\\
\Psi_2(u_2) &
&\begin{turn}{45}\vdots\end{turn}  & \Psi _1(u_1)
\vspace{2mm}\\
\vdots &\begin{turn}{45}\vdots\end{turn}  &
\begin{turn}{45}\vdots\end{turn}  & \vdots \vspace{2mm}\\
 \Psi _N(u_N) &   \Psi _1(u_1)
& \cdots &  \Psi _{N-1}(u_{N-1})
\ea 
\right]Bu \ee                                       
is a gradient vector, 
where $\Psi _i(u_i)=\Phi _i^\dagger (u_i)=\part ^{-1}M_i(u_i),
\,1\le i\le N.$
That is true, indeed. Actually we have
\bea  f_1&=&\Psi f_0= \frac {\delta \tilde H_1} {\delta u},\
{ \tilde H_1} =\int H_1dx,\nonumber \\
H_1&=&(Bu)^T
\left [
\ba {cccc}
\Theta _1(u_1) & \Theta  _2(u_2) & \cdots & \Theta  _N(u_N)
\vspace{2mm}\\
\Theta _2(u_2) &
&\begin{turn}{45}\vdots\end{turn}  & \Theta  _1(u_1)
\vspace{2mm}\\
\vdots &\begin{turn}{45}\vdots\end{turn}  &
\begin{turn}{45}\vdots\end{turn}  & \vdots \vspace{2mm}\\
 \Theta  _N(u_N) &   \Theta  _1(u_1)
& \cdots &  \Theta  _{N-1}(u_{N-1})
\ea 
\right]Bu\nonumber \\&=& 
\sum_{j,k,l=1}^N(\sum_{i=1}^Nb_{i+l-1}b_{k+j-i})u_l(\frac 12 c_ku_{jxx}
+\frac12 d_ku_j+u_ku_j),
\eea 
where the operators $\Theta _i(u_i),\, 1\le i\le N,$ are given by
\be \Theta _i(u_i)=\frac 12 c_i\part ^2+\frac 12 d_i+\frac 23 u_i
+\frac 23 \part ^{-1}u_i\part ,\ 1\le i\le N, \ee 
and $b_i=b_j$ if $i\equiv j\ (\textrm{mod}\, N)$.
We can also choose the energy form for the functional $\tilde H_1$: 
\be  \tilde H_1=\int H_1dx,\ H_1=   
\sum_{j,k,l=1}^N(\sum_{i=1}^Nb_{i+l-1}b_{k+j-i})(-\frac 12 c_ku_{lx}u_{jx}
+\frac12 d_ku_lu_j+u_lu_ku_j). 
\ee

Therefore according to the Magri scheme \cite{Magri-JMP1978} 
\cite{FokasF-LNC1980} \cite{Olver-GTM1071986}, there exist 
other functionals $\tilde H_n,\ n\ge 2,$ such that $\Psi ^nf_0=
\frac {\delta \tilde H_{n}} {\delta u}
,\ n\ge 2.$ 
All such functionals can be generated by computing 
the following integrals
\be \tilde H_n=\int \int_0^1 <(\Psi ^nf_0)(\lambda u),u>d\lambda\,dx,
\ n\ge 0. \label{gradientformular}\ee
Further we can obtain a hierarchy of 
bi-Hamiltonian equations 
\be u_t=K_n:=(\Phi(u))^{n+1}u_x=
J \frac {\delta \tilde H_{n+1}} {\delta u}=M
 \frac {\delta \tilde H_n} {\delta u} ,\ n\ge 0, \label{thirdckdvs}\ee 
which includes the nonlinear system 
(\ref{firstsystemin3rdckdvhierarchy}) as the first member.
It follows that they 
have infinitely many commutative symmetries $\{K_m\}_0^\infty$ and
infinitely many commutative conserved densities $\{H_m\}_0^\infty$. 
All systems of evolution equations 
can reduce to KdV equations once
$u_j=c_i=d_j=0,\ j\ne 1,$ are selected. Therefore they 
are all $N$-component coupled KdV systems.

Let us now work out a concrete example for 
a choice of $A$ with $a_1=1,\,a_i=0,\,2\le i\le N$.

In this case, we have $ b_1=1,\,b_i=0,\,2\le i\le N$.
Then the first Hamiltonian structure is given by
\begin{equation}
\partial _{t}u_{1}=\partial _{x}\frac{\delta \tilde 
H_{1}}{\delta u_{1}},\ 
\partial _{t}u_{k}=\partial _{x}\frac{\delta \tilde H_{1}}{\delta u_{N+2-k}},
\ \part _x=\frac {\part }{\part x},\ 2\le k\le N.\end{equation}
This first Hamiltonian structure has the momentum:
\begin{equation}
\tilde  P=\tilde H_{0}=\frac{1}{2}\int [u_{1}^{2}+
\sum_{k=2}^N
u_{k}u_{N+2-k} ]dx,
\end{equation}
which is of quadratic form with respect to its $N$ Casimirs
(annigilators of the first Poisson bracket) $\tilde F_{k}=\int u_{k}dx$,
$k=1,\cdots,N$. The conservation law of momentum is
\begin{equation}
\partial _{t}H_{1}=\partial _{x}[
\sum_{k=1}^Nu_{k}
\frac{\delta \tilde H_{1}}{\delta u_{k}}-F ],
\end{equation}
where $\tilde H_{1}=\int H_{1}dx$, 
$\partial _{x}F=
\sum _{k=1}^N
\frac{\delta \tilde H_{1}}{\delta u_{k}}u_{kx}$.
For simplicity in the selection (\ref{phi_iform}) 
we can put $d_{i}=0$, because we can eliminate those
constants in the expressions of $\Phi _i$
by making shifts $u_{i}\rightarrow $ $u_{i}-d_{i}/4$.
At this moment our
coupled KdV system reads as
\begin{equation}
\partial _{t}u_{m}=\partial _{x}[\partial _{x}^{2}(
\sum _{k=1}^m c_{m+1-k}u_{k}+
\sum _{k=m+1}^N
c_{N+m+1-k}u_{k})+3
\sum _{k=1}^m u_{k}u_{m+1-k}+3
\sum _{k=m+1}^N u_{k}u_{N+m+1-k}],
\end{equation}
which has the following Hamiltonian
for the first Hamiltonian structure
\begin{equation}
\tilde H_{1}=\int H_{1}dx
\end{equation}
\bea 
H_{1}&=&-\frac{1}{2}[c_{1}w_{1}^{2}+w_{1}
\sum _{k=2}^N
c_{k}w_{N+2-k}+
\sum _{m=1}^{N-1} c_{m}
\sum _{k=2}^{N+1-m}w_{k}w_{N+3-m-k} \nonumber \\ &&
+
\sum _{m=3}^N c_{m}
\sum _{k=N+3-m}^N
w_{k}w_{2N-m+3-k}]+u_{1}^{3}+3u_{1}
\sum_{k=2}^N u_{k}u_{N+2-k} \nonumber \\ &&
+\sum_{k=2}^{N-1}u_{k}
\sum_{m=2}^{N+1-k}u_{m}u_{N+3-k-m}+
\sum_{m=0}^{N-3}u_{m+3}
\sum_{k=0}^m
u_{N-k}u_{N+k-m},\nonumber
\eea 
where $w_{k}=\partial _{x}u_{k},\,
1\le k\le N$, and $N$ is assumed to be greater than two for producing
nontrivial systems.
The second Hamiltonian structure can be determined by its recursion
operator
\begin{equation}
\partial _{t}u_{i}=
\sum_{k=1}^i \Phi 
_{i+1-k}\partial _{x}\frac{\delta \tilde 
H_{0}}{\delta u_{k}}+
\sum_{k=i+1}^N {\Phi }_{N+1+i-k}\partial _{x}\frac{\delta
\tilde H_{0}}{\delta u_{k}},\ 1\le i\le N, \label{exampleckdvs}
\end{equation}
where ${\Phi }_{i}=c_{i}\partial
_{x}^{2}+2(2u_{i}+w_{i}\partial _{x}^{-1}),$ $1\le i\le N$.
The second Hamiltonian structure has the momentum
$F'_{1}=\int u_{1}dx$ and the Hamiltonian $\tilde H_{0}$.

The relationship between the gradients of $\tilde H_{k}$ and $\tilde H_{k+1}$
determined by
recursion operator yields a possibility for constructing an infinite set
of conservation laws and commuting flows by iterations. 
At each step we need to compute integrals to
construct $\tilde H_{k+1}$. They can be done by using the formula
(\ref{gradientformular}) in variational analysis.

Moreover we have
an alternative way to construct an infinite set
of conservation laws and commuting flows.
Let us introduce an eigenfunction
problem for the recursion operator
\begin{equation}
[{\Phi }(u)-\lambda ]{u}_{\tau }=0,
\end{equation}
\begin{equation}
\textrm{i.e.}\ 
\sum_{k=1}^i {\Phi }_{i+1-k}\partial
_{\tau }u_{k}+
\sum_{k=i+1}^N {\Phi }%
_{N+1+i-k}\partial _{\tau }u_{k}=\lambda \partial _{\tau }u_{i},\
1\le i\le N.
\end{equation}
This eigenfunction problem can be equivalently rewritten as
\[ \left\{\ba {l}
\partial _{\tau }u_{k}=\partial _{x}v _{k},
\ 1\le k\le N, \vspace{2mm} \\
\displaystyle { \sum_{k=1}^i {\Phi }_{i+1-k}\partial
_{x}v _{k}+
\sum_{k=i+1}^N {\Phi 
}_{N+1+i-k}\partial _{x}v _{k}=\lambda \partial _{x}v _{i}, \ 1\le i\le N.} 
 \ea \right.
\]
If we use formal series near $\lambda \rightarrow \infty $
\bea 
\partial _{\tau }&=&\partial _{t_{0}}+\frac{1}{\lambda }\partial _{t_{1}}+%
\frac{1}{\lambda ^{2}}\partial _{t_{2}}+\cdots ,
\\ 
v _{i}&=&v _{i}^{(0)}+\frac{1}{\lambda }v _{i}^{(1)}+%
\frac{1}{\lambda ^{2}}v _{i}^{(2)}+\cdots ,
\eea 
we can obtain directly from the above eigenfunction problem
an infinite set of commuting flows.
Furthermore we can use formal series near $\lambda \rightarrow 0$.
Then the first
resulting
nontrivial commuting flow is a coupled long wave equation
\begin{equation}
\sum_{k=1}^i
{\Phi }_{i+1-k}\partial
_{\tau }u_{k}+
\sum_{k=i+1}^N {\Phi }%
_{N+1+i-k}\partial _{\tau }u_{k}=0, \ 1\le i\le N.
\end{equation}
This is an N-component generalization of long wave equation 
(see \cite{DoddEGM-Book1982} for more information about long wave equation),
which commutes with the KdV equation.

\section{Concluding remarks}
\setcounter{equation}{0}

We have proposed new coupled KdV systems possessing 
bi-Hamiltonian structures by extending Hamiltonian operators 
from lower order to higher order of matrix. 
Clearly we may make other choices of $J_i(u_i)$ to give more results
based on Theorem \ref{modelofbigJ}.

Compared to the well-known
coupled KdV systems (for example, see \cite{FordyA-PD1987})
\be  u_t=\Phi ^n u_x,\ 
\Phi (u)= \left[\ba {cccc}
0&\cdots&0&  \Phi _1\vspace{2mm}
\\ 1&\cdots&0& \Phi _2\vspace{2mm}\\
 \vdots& \ddots & \vdots& \vdots\vspace{2mm}\\
0&\cdots& 1& \Phi _N\vspace{2mm}\\
  \ea \right] 
,\ n\ge 0,\label{firstckdvs}\ee 
and quite new couple KdV systems introduced in \cite{Ma-ma82}
\be  u_t=\Phi ^n u_x,\ \Phi(u)= \left [ \ba {cccc} b _N \Phi _1& 
 & & 0 \vspace{2mm}\\
b _{N-1}\Phi _1 +b _{N}\Phi _2& b _N \Phi _1& & \vspace{2mm}\\
\vdots  &\ddots &\ddots &  \vspace{2mm}\\
b _1\Phi _1+\cdots +b _N\Phi _N&
\cdots & b _{N-1}\Phi _1 +b _{N}\Phi _2&b _N\Phi _1
 \ea  \right]
,\ n\ge 0, \label{secondckdvs} \ee
where $u=(u_1,u_2,\cdots, u_N)^T$, $b_i, \, 1\le i\le N,$ are arbitrary 
constants except $b_N\ne 0$,
$\Phi_i=\Phi _i(u_i),\, 1\le i\le N,$ are still defined by (\ref{phi_iform}),
our new coupled KdV systems (\ref{thirdckdvs})
have similar nice integrable properties,
for example, bi-Hamiltonian structures, 
dispersionless limits having bi-Hamiltonian structures (see 
\cite{FerapontovP-PD1991}
for the case of the well-known coupled KdV systems).
But there exist differences among the structures of 
recursion operators corresponding to 
these three hierarchies of coupled KdV systems.
The bi-Hamiltonian structure (see \cite{Ma-ma82}) of 
the hierarchy (\ref{secondckdvs})
can similarly be derived from the Hamiltonian operators formed by
(\ref{form1concrete1}) 
in the first example of the second section.
This is why we didn't deliver above a detailed analysis
for constructing integrable systems starting from (\ref{form1concrete1})
or equivalently from (\ref{form1concrete2}).
However
we don't know whether or not two new coupled KdV hierarchies 
(\ref{secondckdvs}) and (\ref{thirdckdvs})
have other integrable properties, for example, 
Lax pairs like (\ref{firstckdvs}).
 
In terms of the existence of recursion operators, 
other integrable couple KdV systems, say, Jordan KdV systems,
have also been derived 
(see for example
\cite{Svinolupov-PLA1989}
\cite{FokasL-PRL1996} \cite{Karasu-IJTP1997} 
\cite{GursesK-solv-int9711015}).
A natural question is 
whether there exist other hierarchies possessing 
bi-Hamiltonian structures among those coupled KdV systems.
This will enrich the content of Hamiltonian theory for coupled KdV systems. 
 
\vskip 3mm
\noindent{\bf Acknowledgments:} The authors
would like to thank 
City University of Hong Kong for financial support.
One of the authors (W. X. Ma) is also grateful to Prof. Y. S. Li and 
Prof. F. K. Guo for valuable discussions.


\newpage
\small 
\baselineskip 13pt
\begin{thebibliography}{99}
\bibitem{GelfandD-FAA1979}I. M. Gel'fand and I. Y. Dorfman,
Funct. Anal. Appl. 13 (1979) 248.
\bibitem{FuchssteinerF-PD1981}B. Fuchssteiner and A. S. Fokas, 
Physica D 4 (1981) 47.
\bibitem{Magri-JMP1978}F. Magri, 
   J. Math. Phys. 19 (1978) 1156.
in: {\it Lectures Notes in Physics} Vol. 120 (Springer-Verlag, 
Berlin, 1980) pp233.
\bibitem{FokasF-LNC1980}A. S. Fokas and B. Fuchssteiner,
  Lett. Nuovo Cimento 28 (1980) 299.
\bibitem{Olver-GTM1071986}P. J. Olver, {\it Applications of Lie Groups 
to Differential Equations} (Springer-Verlag, New York, 1986).
\bibitem{Ma-GJ1986}W. X. Ma, 
  J. Graduate School, USTC and Academia Sinica 3(3) (1986) 37.
\bibitem{TuM-JPDE1992}G. Z. Tu and W. X. Ma, 
J. Partial Diff. Equ. 3 (1992) 53.
\bibitem{GhoshC-JPSJ1994}C. Ghosh and A. R. Chowdhury,
  J. Phys. Soc. Jpn. 63 (1994) 3911.
\bibitem{MaF-PLA1996}W. X. Ma and B. Fuchssteiner,
  Phys. Lett. A 213 (1996) 49.
\bibitem{MaF-CSF1996}W. X. Ma and B. Fuchssteiner,
  Chaos, Solitons $\&$ Fractals 7 (1996) 1227.
\bibitem{Ma-JPA1998}W. X. Ma, Extension of hereditary symmetry operators,
solv-int/9803002 (1998).
\bibitem{DoddEGM-Book1982}R. K. Dodd, J. C. Eilbeck, J. D. Gibbon and
H. C. Morris, {\it Solitons and Nonlinear Wave Equations},
(Academic Press, London, 1982).
\bibitem{FordyA-PD1987}A. P. Fordy and M. Antonowicz, 
Physica D  28 (1987) 345.
\bibitem{Ma-ma82}W. X. Ma,
  ``Generalized KdV systems and their bi-Hamiltonian formulations,'' 
  solv-int/9803009 (1998).
\bibitem{FerapontovP-PD1991} E. V. Ferapontov
and M. V. Pavlov, 
Physica D 52 (1991) 211.
\bibitem{Svinolupov-PLA1989}S. I. Svinolupov,
 Phys. Lett. A 135 (1989) 32.
\bibitem{FokasL-PRL1996}A. S. Fokas and Q. M. Li,
 Phys. Rev. Lett. 77 (1996) 2347.
\bibitem{Karasu-IJTP1997}A. Karasu,  Intern. J. Theoret. Phys. 36 (1997) 705.
\bibitem{GursesK-solv-int9711015}M. G\"urses and A. Karasu,
  ``Integrable coupled KdV systems,'' 
  solv-int/9711015 (1997).
\end{thebibliography}
\end{document}